\documentclass[aps,prx,twocolumn,showpacs]{revtex4-1}

\usepackage{amsmath}
\usepackage{amssymb}
\usepackage{dcolumn}
\usepackage{bm}
\usepackage{hyperref} 

\usepackage[english,russian]{babel}
\usepackage[utf8]{inputenc}

\newcommand{\wh}{\widehat} 
\newcommand{\wt}{\widetilde}
\newcommand{\mc}{\mathcal} 

\newcommand{\pr}{^\prime}

\newcommand{\pa}{\partial}
\newcommand{\vro}{\varrho}
\newcommand{\ep}{\epsilon} 
\newcommand{\vep}{\varepsilon} 
\newcommand{\vd}{\varDelta}
\newcommand{\vk}{\varkappa}
\newcommand{\lm}{\lambda}

\newcommand{\df}{\mathcal{D}}
\newcommand{\ef}{\mathcal{E}} 

\newcommand{\dof}{\widehat{\mathcal{D}}} 
\newcommand{\jof}{\widehat{\mathcal{J}}}
\newcommand{\kof}{\widehat{\mathcal{K}}}  
\newcommand{\lof}{\widehat{\mathcal{L}}}
\newcommand{\nof}{\widehat{\mathcal{N}}} 

\newcommand{\qof}{\widehat{\mathcal{Q}}} 

\newcommand{\sof}{\widehat{\mathcal{S}}}

\newcommand{\qtf}{\widetilde{\mathcal{Q}}}
\newcommand{\stf}{\widetilde{\mathcal{S}}}

\begin{document}

%\Russian

\title{Kinetics, Pseudo-Kinetics, Uncertainty Principle and Quantum 1/f Noise}

 \author{Yu.\,E.\,Kuzovlev}
% \affiliation{Donetsk Institute for Physics and Engineering}
% \email{kuzovlev@kinetic.ac.donetsk.ua}
 \affiliation{Donetsk Free Statistical Physics Laboratory}
 \email{yuk-137@yandex.ru, kuzovlev@fti.dn.ua}

%\date{March 2018}

\begin{abstract} 
1/f noise at arbitrary low frequences 
is the way of existence of irreversibility 
in thermal motion governed by reversible laws of mechanics. 
This statement not once was confirmed in statistical mechanics 
beyond its traditional kinetical roughenings.  
Here we point out that in case of quantum statistical mechanics 
in principle it is sufficient to avoid such the roughening 
as the ``Fermi golden rule''.    
This means taking into account the time-energy uncertainty principle 
(time-frequency one in classical limit) and thus uncertainties 
in characteristics of real collisions and scatterings of  
particles and/or quanta. 
We consider the resulting ``pseudo-kinetics''  and 
demonstrate how it  produces quantum 1/f-noise.  
\end{abstract}


\maketitle

%------------

%\tableofcontents

%%%%%%%%%%%%%%%%%%%%%%%%%%%%%%%%%%%

% \epigraph{\it The greatest difficulties lie where we are not looking for them} {J. W. von Goethe}

\section{Introduction}

{\bf 1}.\, Not long ago in \cite{lufn} we have reviewed 
the idea for the first time pronounced many years ago 
in \cite{p157,bk1,bk2,ufn}:\, any physical random process 
gives a place for 1/f noise. 
Indeed, mere logics prompts, for instance, that 
if a process produces random events but constantly forgets 
their previous numbers then it is definitely unable 
to control their production rate at large time intervals, 
hence, producing also arbitrary long deviations of the rate 
from its past values, that is flicker type rate fluctuations. 

In classical statistical mechanics (SM) 
such the way of thinking gets support \cite{99,i1,97,tmf,pufn,gs}
when one can spy upon correlations due to determinism 
of mechanical motion and respectively discard statistical hypotheses 
principally superfluous in view of the determinism. 
The forgetting of the past here is the same as complexity of 
mechanical motion. Because of its time reversibility 
its future also gets forgotten, 
in the sense ot its unpredictability, 
thus acquiring features of irreversibility: 
relaxation, dissipation, diffusion and noise.
At that, rates of relaxation, dissipation, diffusion and noise  
are not attributes of mechanics' laws 
and therefore can appear from these laws 
in the form of fluctuations only. 

Moreover, these fluctuations alwaays are of flicker type. 
Indeed, since the rates are time-non-local 
characteristics, their fluctuations say about 
prehistory of present  system's state  
and therefore have no influence on the future 
as fully determined by present  instant state. 
Hence, the rates' fluctuations do not cause a feedback reaction.
Hence, they they have no definite relaxation times.   

We see that 1/f-noise arises as another side of usual 
(``white''') thermal noise and hides no specific physical 
``mechanisms''. Instead of them, its theory needs in adequate 
mathematical approaches to SM equations. So, the story  
of 1/f-noise strongly resembles the ancient story 
of the phlogiston which eventually was understood as 
thermal agitation of particles of the matter.

{\bf 2}.\, In quantum SM both the past
and the future are in a fog of quantum uncertainty, 
so that connection between them is not fatal 
and allows some ``free will''. 
At the same time the quantum uncertainty obeys 
completely deterministic laws of evolution, 
and therefjre all the aforesaid about 1/f noise 
covers quantum case as well. 

This was demonstrated for 
tunnel transport noise in many-electron systems \cite{kmg} 
and 
quantum random walk of a particle interacting with 
thermal bath \cite{p1207}, in particular, electron in 
phonon bath (thermally oscillating  crystal lattice) 
\cite{oct,eph}. 
These investigations confirmed title of the work \cite{hv},  
``Lattice scattering causes 1/f noise'',   
and 
showed how it arises just from electron-lattice 
(particle-bath) interaction by itself, 
without any external driving. 

The above claimed verbal proof of inevitability 
of 1/f-noise in irreversible phenomena was   
formally  fortified by proving theorems  
\cite{p1207,oct} 
which do connect statistics of particle's random walk in 
bath medium to exact density matrix evolution of the system 
and state  
that diffusivity of the particle must possess   
flicker fluctuations because their absence is 
in contradiction to unitarity of the evolution in SM. 

Such principal results stimulate search for 
more light approximate methods simplifying SM as much 
as possible but without loss of 1/f-noise, 
in contrast to usual kinetics always losing it. 
Interestingly, this requirement can be satisfied 
if one takes into account factual duration 
(of quantum amplitude formation) of quantum transitions. 
The works \cite{kmg} and \cite{eph} gave clear illustrations 
of usefulness of this recipe, 
and this aspect of the theory will be focus 
of our consideration below.

\section{1/f-noise as surprise of theoretical physics} 

\subsection{The golden rule of kinetics}
 
{\bf 1}.\,  The word ``surprise''  is taken from the 
R.Peierls' book \cite{sur} 
and describes typical relations between 
theoreticians and theories in physics 
under usual mix of findings and mistakes.  
Theoreticians hope to meet pleasant surprises, 
as is indicated by W.Pauli's ``law of conservation of sloppiness'' 
cited in \cite{sur}. 

We will dispute this optimism at least in the case discussed 
in section 5.3 in \cite{sur}, namely,  the question of applicability 
of the famous ``Fermi golden rule'' \cite{vk,hj}.
The golden rule (GR) is usual artificial addition to 
quantum-mechanical non-stationary perturbation theory (QNPT) 
in order to obtain kinetic equations, particularly, 
for charge carriers in solids \cite{hj,lp}. 

Let us recall general expression   
\begin{eqnarray} \label{vp0}
p_\tau(2\leftarrow 1) =  
\frac { 2\,[1 -\cos\,( (\mathcal{E}_2 - \mathcal{E}_1)\, \tau/\hbar)] } 
{ (\mathcal{E}_2 - \mathcal{E}_1)^2 } \, |\Phi_{21} |^2 \,\, 
\end{eqnarray}
of the lowest order of QNTP  \cite{ll3} for probability 
(square of modulus of quantum amplitude) 
of that during time \,$\tau$\, 
under influence of perturbation  \,$\Phi $\, 
with matrix elements 
 \,$\Phi_{21} = \langle 2|\Phi |1\rangle$\, 
 a quantum system makes trensition 
 between stationary (in absence of the perturbation) 
 states  \,$|1\rangle$\, and \,$|2\rangle$\,   
with energies \,$\mathcal{E}_1$\, and \,$\mathcal{E}_2$\,. 
It will be comfortable to write (\ref{vp0}) as 
\begin{eqnarray} \label{vp}
p_\tau(2\leftarrow 1) =  \tau\, \gamma_\tau(2\leftarrow 1)\,\, , \,\,\,
\\  \nonumber
\gamma_\tau(2\leftarrow 1) =
\frac { 2\pi }\hbar\, |\Phi_{21} |^2 \, 
\delta_{1/\tau} (\mathcal{E}_2 - \mathcal{E}_1) \,\, \, 
\end{eqnarray}
with probability of transition per unit of time of transition 
,$\gamma_\tau(2\leftarrow 1)$\, 
and 
normalized to unit sharp peak-like 
quasi-delta function (QDF) 
\begin{eqnarray} \label{kdf}
\delta_{1/\tau}(\epsilon) \equiv \frac \hbar{2\pi \tau}\, 
\left[ \frac { 2\,\sin\, (\epsilon \tau/2\hbar)} \epsilon \right]^2 \,\,  \,\, 
\end{eqnarray} 
which turns into usual Dirac's delta-function (DDF) 
in the limit  
 \,$ \tau\rightarrow\infty$\,:\,\,$\delta_0(\epsilon)=\delta(\epsilon)$\,.  

Formally the GR is performing in (\ref{vp}) 
the QDF replacement with DDF, 
\begin{eqnarray}  \label{sur} 
\delta_{1/\tau}(\epsilon) \,\Rightarrow \, \delta(\epsilon) \,\, ,\,\,   
\end{eqnarray}
under assumption that energies of the states belong to  continuous spectrum. 
Or, what is the same, replacement  
\begin{eqnarray} \label{g0}
\gamma_\tau(2\leftarrow 1)  \,\Rightarrow\,  
   \gamma_\infty(2\leftarrow 1)  \,\, , \,
\end{eqnarray} 
which enforces time-dependent ``probabilities per unit time'' 
to become constants independent on duration of transitions.   

{\bf 2}.\, If full Hamiltonian of system under consideration is 
\begin{equation} \label{h0}
H=  H_0 + \Phi \,\, ,  \,\, 
\end{equation} 
then \,$|1\rangle$\, and \,$|2\rangle$\, 
can be thought eigen-states of \,$H_0 $\,:\, 
\,$H_0 |1\rangle =\ef_1 |1\rangle$\, and so on.  
For definiteness we may keep in mind the 
``Brownian'' particle (BP) in a medium  (``thermostat'') 
with Hamiltonian (9) from \cite{lufn},  with 
\begin{equation} \label{h00} 
H_0 =  \frac {P^2}{2M} + H_{th} \,\,  \,\, 
\end{equation}
representing system without interaction between its parts 
(BP and medium) while \,$\Phi $\, represents their interaction. 

The kinetic theory treats quantum evolution 
like  jump-like transitions between states of a given set 
as if 
next time jumps realize or not dependently on 
outcomes of imaginary chance trials
with chances established from above 
(``dice tossings'' by god of mathematicians). 
Following this concept, dividing (long enough) 
time of system's evolution into (not too long) 
intervals with duration \,$\tau$\,, 
each almost surely containing not more then one transition, 
and looking through all such series of transitions 
``by the probability theory'' \cite{fel}, 
one can conclude 
that (density of) probability \,$\rho$\, 
of finding our system to be in one or another state 
(eigen-state of unperturbed Hamiltonian) 
is given by 
\begin{eqnarray} \label{evt}
\rho(t) \approx (1 + \tau \widehat{\mathcal{K}}_{1/\tau})^{\,t/\tau} \,\rho(0) 
\approx \exp{(t\kof_{1/\tau})} \,\rho(0)  \, \, \,
\end{eqnarray} 
with kinetic operator (KO) \,$\widehat{\mathcal{K}}_{1/\tau}$\,
which acts by formula
\begin{eqnarray} \label{kot} 
(\widehat{\mathcal{K}}_{1/\tau}\,\rho)_1  \equiv \sum_2 \,[ \gamma_\tau(1\leftarrow 2) \, \rho_2 
- \gamma_\tau(2\leftarrow 1) \, \rho_1 ]    \,\, \,    
\end{eqnarray}
with ciphers in the role of indices of possible states.

Such the theory is very uncomfortable because of arbitrariness 
of the quantity \,$\tau$\, about which one only can say that  
it certainly is bounded both from above and from below. 
Kinetics disposes of this inconvenience, 
like frequently in mathematical physics, 
by means of physically absurd idealization 
when 
the factor \,$\tau$\, before 
the KO symbol and in the exponent (number of intervals \,$t/\tau$\,) 
in  (\ref{evt}) is tended to zero 
while inside the KO (in QDF) the same factor is tended to infinity 
As the result of such wild arbitrariness, 
one creates formally unambiguous evolution law 
\begin{eqnarray} \label{ev0}
\rho(t) =  \exp{(t\kof_0)} \,\rho(0)  \, \, \,
\end{eqnarray} 
or, in differential form, 
\begin{eqnarray} \label{ku0}
\frac {\partial \rho}{\partial t} = \widehat{\mathcal{K}} _0\,\rho\, \,  \,\,
\end{eqnarray}
with KO  \,$\widehat{\mathcal{K}}_0$\, acting by formula   
\begin{eqnarray} \label{ko0} 
(\widehat{\mathcal{K}}_0\,\rho)_1 =  \sum_2 \,[ \gamma_\infty(1\leftarrow 2) \, \rho_2 
- \gamma_\infty(2\leftarrow 1) \, \rho_1 ]    \,\,    
\end{eqnarray}
with the sum symbol presuming at once integration 
over continuosly varying indices 
and summation over discrete ones. 

Equation (\ref{ku0}) with right-hand side  constructed like (\ref{ko0}) 
is termed ``master equation'' \cite{vk}. 
In the context of quantum theory an equation of such kind 
for the first time was suggested by Pauli \cite{pa} already in 1928. 
At that, Pauli considered transitions between not individual states 
but groups of states with close energies, 
assuming compldete randomness of their phases and applying averaging over them. 
This trick eventually is equivalent to application of GR. 

Unfortunately, kinetic equations are so much extremal roughenings of SM 
that they categorically lose flicker noise. 
In order  to make sure of this, first let us compare defects 
of the approximations (\ref{evt}) and (\ref{ev0}) 
to recognize  another approximation which is able to catch flicker noise.

\subsection{Beyond the golden rule: \\ failure of kinetics 
and advances of pseudo-kinetics}
  
{\bf 1}.\, From viewpoint  of quantum mechanics, 
the golden rule approximation (GRA) (\ref{ev0})-(\ref{ko0}) 
seems more rough than (\ref{evt})-(\ref{kot}) 
but instead much more simple. 
Therefore the attempt in \cite{sur} to justify just the GRA   
by comparing KO (\ref{kot}) and KO (\ref{ko0})  
looks very natural.

Of course, these KOs differ  to the extent of narrowness 
of the QDF \,$\delta_{1/\tau}(\epsilon)$\,'s peak, 
i.e. smalness of QDF's width  
\,$ \varDelta\epsilon \approx 2\pi\hbar/\tau$\,,
in comparison with width \,$\epsilon_0$\, 
of factor  \,$|\Phi_{21} |^2\rho_2$\, 
as function of difference \,$\mathcal{E}_2 - \mathcal{E}_1$\,. 
 According to the reasonings exposed in \cite{sur} 
 in case of interactions - collisions - of a free charge carrier 
 (let electron) with phonons and defects of crystal lattice  
 in semiconductors  and even metals  practically always    
 \,$\varDelta\epsilon/\epsilon_0 \ll 1$\,, 
 even if assuming that \,$\tau$\, is less than mean time between 
 collisions, \,$\tau_0$\,. 
This statement serves in \cite{sur} as base for conclusion  
that the ``sloppiness'' of GRA is negligible, thus excusable,
i.e.  GRA has no significant alternatives. 

From our viewpoin, however, that is wrong coclusion, 
for one should compare not KOs  (\ref{ko0}) and (\ref{kot}) in themselves 
but corresponding evolution operators, 
that is exponentials in (\ref{ev0}) and (\ref{evt}).
All the more because these operators have qualitative differences, 
at least in their relations to ``unpertubed energy'' \,$\mathcal{E}$\,. 

{\bf 2}.\, Namely, evolution due to (\ref{evt}) is accompanied 
with unbounded spreading, or diffusion, of the distribution \,$\rho(t)$\, 
over the energy \,$\mathcal{E}$\,.
It is so merely because 
at any fixed  \,$\tau$\,'s value  
each particular transition gives randon increase or decrease 
of \,$\mathcal{E}$\, by a step of order of QDF's width 
\,$\sim \varDelta\epsilon \approx 2\pi\hbar/\tau$\,,
and these steps (as well as successive transitions themselves)  
are mutually statistically independent 
(incoherent).  
It is in obvious contradiction to (quantum-mechanical) reality. 

In opposite, evolution under GRA (\ref{ev0}) due to \,$\varDelta\epsilon =0$\, 
precisely conserves the unperturbed energy \,$\mathcal{E}$\, 
and thus all time passes not leaving some hyper-surface 
\,$\ef =$\,const\, in the space of indices  
(\,$H_0 =$\,const\, in the space of states). 
This behavior also is in contradiction to physical reality, 
since in fact  \,$H_0$\, is not integral of motion. 

Thus, kinetics in any variant is physically inadequate theory. 
Fortunately, our comparison of evolutions with \,$\tau =$\,const\,$<\infty$\, 
and  \,$\tau =\infty$\, prompts third, original, possibility  
when duration of ripening of transitions \,$\tau$\,  
is not fixed but grows together with duration of system's evolution 
and observation \,$t$\, and takes the only finite value which is free of 
theoretical arbitrariness, that is \,$\tau =t$\,, 
so that 
\begin{eqnarray} \label{ev}
\rho(t) = \exp{(t \kof_{1/t})} \,\rho(0) \, \, . \, 
\end{eqnarray}
In such way one removes fundamental defect of GRA, 
i.e. replacing of actual transitions, 
which are squeezed in time inside frames of system 
observations, by fantastic transitions stretched over infinite time.
Formally the value  \,$\tau =t$\, lies between \,$\tau =$\,const\, 
fnd  \,$\tau =\infty$\,, therefore the corresponding evolution law  
(\ref{ev}) can be named ``golden mean  approximation'' (GMA). 

Clearly, evolution in GMA does not conserve the unperturbed energy 
\,$\mathcal{E}$\,. But its probable departure from its initial value 
all the time keeps closed between nearly constant finite bounds, 
since QDF's width \,$\sim \varDelta\epsilon \approx 2\pi\hbar/t$\, 
and thus magnitude of steps of random walk over \,$\ef$\, 
do decrease as evolution (observation) time increases. 
This is physically plausible picture which makes GMA 
reasonable alternative to GRA and thus surprise in the sense of \cite{sur}.

At the same time, evidently, this picture no more belongs to the kinetics,
because now ``transition probabilities (per unit time)'' 
have no definite values. Instead, they become very significantly varying 
with evolution time, so that, in essence, they lose naive probabilistic  
meaning at all. 
In this way the theory gives back at least a part 
of coherence of (quantum-) mechanical motion. 
Therfore such approximations can be named also ``pseudo-kinetic'', 
while any  operator like \,$\kof_{1/t}$\,, 
containing QDF  \,$\delta_{1/t}(\epsilon)$\,,  
``pseudo-kinetic operator'' (PKO). 

Importantly, the PKO automatically introduces unrestrictedly
 low frequences \,$\sim 1/t$\, as mathematical building material 
for flicker noise. Next, consider how principally it happens. 

{\bf 3}\, Now let us compare KO  \,$\widehat{\mathcal{K}}_0$\, 
and PKO  \,$\widehat{\mathcal{K}}_{1/t}$\, from viewpoint 
of their stationary probability distributions \,$\rho_{st}$\,,  
i.e. solutions to equations \,$\widehat{\mathcal{K}}_0\rho =0$\, 
andЩ \,$\widehat{\mathcal{K}}_{1/t}\,\rho =0$\,.    
Since KO conserves the unperturbed energy, it possesses infinitely many 
linearly independent stationary distributions, for instance, 
microcanonical \,$\rho_{st} \propto \delta(\mathcal{E}-\mathcal{E}_0)$\,, 
that is uniform on some given hyper-surface 
\,$\mathcal{E}=\mathcal{E}_0=$\,const\, in index space. 
In opposite, since PKO does is not conserving the energy,  
it has only one  stationary distribution  \,$\rho_{st} =\,$const\,, 
uniform in the whole index space. 
 
In the language of spectral theory of linear operators, 
both KO and PKO have zero eigen-value, but in case of KO it is 
infinitely degenerated while in case of PKO non-degenerated. 

In other words, change of KO to PKO fully kills the degeenration. 
Taking into account essential cognation of KO and PKO, 
at \,$\varDelta\epsilon/\epsilon_0 \approx 2\pi\hbar/\epsilon_0 t \ll 1$\, 
we can consider this change as weak perturbation of KO 
in the sense of the abstract perturbation theory (PT) \cite{kato}. 
If so, the removal of degeneration of zero eigen-value (EV) 
means that the corresponding eigen-functions (EF),  
after transforming under perturbation, do acquire new EVs
which (all except one) are non-zero though small to the extent 
of perturbation parameter \,$2\pi\hbar/\epsilon_0 t\propto 1/t$\,. 

Here just such CVs slipping to zero  \,$\propto 1/t$\, are responsible 
for 1/f-noise, for they imply presence of arbitrary slow relaxation modes 
in evolution by PKO and thus arbitrary long statistical correlations. 
Namely, that are modes connected to the additional (activated)  
degree of freedom along \,$\ef$\,. In \cite{eph} we presented 
the example: fourth-order correlations of electron velocity  
(in phonon medium) which represent 1/f-noise in electron's 
 rate of diffusion (diffusivity). 

In principle, there is nothing except trivial uncertainty 
in energy transfer under transitions 
(or in precision of resonances in the classical limit) 
due to finiteness of interaction duration.  
 Next, consider it in more detail.

\section{From uncertainty principle to 1/f interaction fluctuations}

{\bf 1}.\, Preliminarily, notice that because of symmetry 
\,$\gamma_\tau(2\leftarrow 1)=$\,$\gamma_\tau(1\leftarrow 2)$\, 
PKO is symmetric operator. Therefore all its EVs are real 
and its action can be represented in the form 
\begin{eqnarray} \label{pko-} 
(\widehat{\mathcal{K}}_{1/t}\,f)_1  = \sum_2 \, 
\gamma_t(1\leftarrow 2)\, [\, f_2 - f_1 \,] \,\, . \,\,    
\end{eqnarray}
From here for any function \,$f$\, of state indices 
one can easy derive inequality
\begin{eqnarray} \nonumber    %  \label{lz} 
 \sum_1 f_1^* (\kof_{1/t} f)_1  
 = - \frac 12 \sum_{1,2}  \gamma_t(1\leftarrow 2) \,
 |f_1-f_2|^2\, \,\leq \, 0 \,\, , \,  
\end{eqnarray} 
and from here deduce (physically doubtless) statement that   
all EVs of PKO are non-positive.   

{\bf 2}.\, Let us focus at random walk, or fluctuations, 
of unperturbed energy  \,$\ef$\,. 
When taken with opposite sign, these fluctuations can be interpreted, 
in view of conservation of the total energy $H_0+\Phi$\,, 
also as fluctuations of interaction energy. 
We are interested in their marginal probability distribution 
\begin{eqnarray} \label{re} 
W(\ef) =  \sum_1 \,\delta(\ef -\ef_1)\,\rho_1\, \, , \,\,    
\end{eqnarray}
its time evolution and their temporal correlations.
For simplicity let us assume that conditions of \,$\ef$\,'s walk 
are uniform in the state space:\, \,$\rho$\, is uniform 
on any equal \,$\ef$\, hyper-surface, and quantity  
\begin{eqnarray} \label{fe} 
g(\ef -\ef_2) \equiv 
 \sum_{1} \,\delta(\ef -\ef_1)\, \frac {2\pi}\hbar\,
 |\Phi_{12}|^2 \,\,  \,    
\end{eqnarray}
is a function of only difference of energies before and after 
transitions. Evidently, it then is non-negative and even function:\, 
\,$g(-\ep) =g(\ep) \geq 0$\,. 
Then PKO reduces to an operator acting in the set  
of functions \,$f(\ef)$\,  of one variable \,$\ef$\, by formula   
\begin{eqnarray} \label{koe} 
\kof_{1/t} f(\ef) = \int \delta_{1/t} (\ef-\ef^\prime)\, 
\, \times\,\,\, \\   \nonumber   \times\,\, 
g(\ef -\ef^\prime)\,\,[\, f(\ef^\prime) - f(\ef)]\,\, d\ef^\prime \,\, .  \,\,    
\end{eqnarray}
This operator possesse clear continuous spectrum of EFs 
\,$(2\pi\hbar)^{-1/2}  \exp{(i\theta\ef/\hbar)}$\,. Their CVs are
\begin{eqnarray} \label{sc} 
\lm_t(\theta) = \int \delta_{1/t} (\ep)\,  
g(\ep)\, [\,\cos{(\theta\ep/\hbar)} -1\,] \,d\ep\, \,  \,\, 
\end{eqnarray}
and all belong to interval \,$-g_t <\lm_t(\theta) \leq 0 $\,  
with lower bound  
\,$-g_t\equiv  -\int \delta_{1/t} (\ep) \,g(\ep)\,d\ep$\,. 
Correspondingly, 
\begin{eqnarray} \label{prope} 
\exp{(t\kof_{1/t} )} \,f(\ef) = \int S_{1/t}(t,\ef-\ef^\prime)\, 
f(\ef^\prime) \, d\ef^\prime \,\,   \,\,    
\end{eqnarray}
with propagator - or, in terms of theory of random processes,  
transient probability (density) of the energy, - what may be expressed by   
\begin{eqnarray} \label{poe} 
S_{1/t}(t^\prime,\ef-\ef^\prime) = \int e^{\,\lm_t(\theta)\,t^\prime}\, 
e^{\,i\theta\,(\ef-\ef^\prime)/\hbar} \, 
\frac {d\theta}{2\pi\hbar}  \,\, .  \,\,    
\end{eqnarray}
Alternatively, by dividing PKO (\ref{koe}) into two obvios parts, 
this propagator can be represented in the form of series: 
\begin{eqnarray} \label{eve} 
S_{1/t}(t^\prime,\ep) = 
e^{-g_t t^\prime} \sum_{n=0}^\infty\, \frac {(g_t t^\prime)^n} {n!} \, 
( D_t(\ep) \,\otimes)^n \,\delta(\ep) \,\, , \,\, \,   
\end{eqnarray}
where 
\,$\ep=\ef-\ef_0$\,, the symbol \,$\otimes$\, denotes operation 
of convolution over variable \,$\ep$\,, 
and we introduced function
\begin{eqnarray} \label{dg} 
D_t(\ep) \equiv \frac { \delta_{1/t} (\ep) \,g(\ep)} 
{ \int \delta_{1/t}(x) \,g(x) \,dx}  = 
\frac { \delta_{1/t}(\ep) \,g(\ep)} {g_t} \,\, . \,\,  
\end{eqnarray}

We want to trace in this model how vanishingly small uncertainties 
\,$\vd\ep\sim 2\pi\hbar/t$\, 
of energy transfers in particular transitions 
do summation into non-vanishing fluctuations of rate of the interaction.

{\bf 3}.\, Let at the beginning of observaion 
the energy is definitely known: \,$W(t=0,\ef)=\delta(\ef-\ef_0)$\,. 
We will consider, in the frameworks of GMA (\ref{ev}) and PKO (\ref{koe}), 
probability distribution 
 \,$W(t,\ef) = \exp{(t\kof_{1/t})} \,W(0,\ef)$\, 
 of following energy fluctuations.

First, calculate their variance by using the  series (\ref{eve}). 
Obviously, it has meaning of expansion over 
number  \,$n$\, of elementary transitions 
and prescribes them Poissonian statistics 
with mean number of transitions \,$\langle n\rangle =g_t t$\,, 
while non-negative, and normalized to unit, function 
(\ref{dg})  plays role of probability distribution of 
 discrepancy (non-conservation, uncertainty) of energy  
 in separate transition. 
At that \,$n$\,-fold convolution of $D(\ep)$\, represents sum of \,$n$\, 
mutually  statistically independent contributions. 
Since variance of such sum equals to sum of variances of the  
contributions, one finds   
\begin{eqnarray}   \label{dis} 
 \int (\ef-\ef_0)^2\,W(t,\ef) \,d\ef  =  
 \langle n\rangle \int \ep^2\,D_t(\ep) \,d\ep = \,\,\, \, 
\\   \nonumber = 
 t \int \ep^2\,\delta_{1/t} (\ep) \,g(\ep) \,d\ep \,\approx\,  
\frac {\hbar}\pi \int g(\ep)\, d\ep \,\, . \,\,  
\end{eqnarray}
At the end here we neglected QDF's oscillations, 
since at large enough time the function \,$g(\ep)$\, 
may be thought much more smooth than \,$\cos{(\ep t/\hbar)}$\,.
As one can see, under  \,$\int g(\ep)\, d\ep <\infty$\, 
the variance  tends with time to a finite value 
which is fully determined by the inversely-quadratic ``wings'', 
or ``long tails'' \,$\propto \ep^{-2}$\,, of QDF, 
that is just by what is discarded in GRA. 

{\bf 4}.\, All the more, the QDF's tails determine statistics 
of energy fluctuations in case when \,$\int g(\ep)\, d\ep =\infty$\, 
and the variance turns into infinity. 
According to (\ref{fe}), this is the case if  
\,$(2\pi/\hbar) \sum_{1} |\Phi_{12}|^2  =\infty$\,, 
i.e. 
the matrix elements (ME) of transitions by themselves, 
without additional requirement of energy conservation, 
allows for infinitely many variants of transitions from 
a given state. This is both theoretically thinkable and physically realistic 
situation. Let us discuss it with representation (\ref{prope})-(\ref{poe}). 

We may start it from simple special case when \,$g(\ep)=g=$\,const\, 
(by themselves ME equally connect states with any energy difference). 
Subjecting QDF to Fourier transform, one has  
\begin{eqnarray} \label{pf} 
\int \delta_{1/t} (\ep)\,  
\cos{(\theta\ep/\hbar)} \,d\ep\, = 
\left[1- \frac {|\theta|}t \right]\,\b1(t- |\theta|) \,\, , \,\,\,\, 
\end{eqnarray}
where
\,$\b1(\cdot)$\, is the Heaviside step function.  
Therefore, if  \,$|\theta|<t$\, then EVs (\ref{sc}) 
are equal to  \,$\lm_t(\theta) =- g|\theta|/t$\,, 
so that the exponentials in (\ref{poe}) 
become time-independent. 
Hence, in the region of not too small energy fluctuations, 
at \,$|\ep|=|\ef-\ef_0|\gtrsim \hbar/t$\,, their distribution takes 
(quasi-)  stationary form 
\begin{eqnarray} \label{rk} 
W(t,\ef) \approx  \int e^{-g|\theta|}\, 
\cos{(\theta\ep/\hbar)} \,\frac {d\theta}{2\pi\hbar} = \,\,\, 
 \\  \nonumber  = \,
 \Delta _{g}(\ep)\, \equiv\, 
 \frac 1\pi\, 
\frac {g\hbar} {g^2 \hbar^2 + \ep^2}  \,\, , \,      
\end{eqnarray} 
that is the form of Cauchy distribution. 

{\bf 5}.\, Now recall the qualitative specificity 
of Cauchy distributions (CD) what differs CD from distributions with 
finite statistical moments (for example, Gaussian ones, 
\,$G_g(\ep) = (\sqrt{2\pi} g\hbar)^{-1}\,\exp{(-\ep^2/2g^2\hbar^2)}$\,). 
Namely, under convolutions of CDs summation of their widths 
takes place:\,   
\,$\Delta_g \otimes \Delta_g = \Delta_{2g}$\,\,,\,  
instead of squares of widths 
(as under convolutions of Gaussians: 
\,$G_g \otimes G_g = G_{\sqrt{2}\,g}$\,\,). 
Such specificity of Cauchy statiteistics may be more brightly 
expressed by saying that mutually independent random quantities 
distributed  ``by Caucgy'' 
are added as if they were completely dependent 
(e.g. literally equal).  
So paradoxical property explains why 
distributions close to CD involuntarily appear in both phenomenological 
and microscopic theories of flicker noise with the origin 
characterized in Introduction \cite{lufn,bk1,bk2,99,tmf,pufn,gs}. 

Formally the mentioned property of CD is due to its 
quadratic tails, and therefore it extents to other distributions 
with such tails, including  our QDF. 
Just by this reason the sum of widths of  
 \,$\langle n \rangle = gt$\, convolutions of QDF 
 in the expansion  (\ref{eve}) 
gives fixed finite width of CD in (\ref{rk}). 
It shows that under neglect of QDF's oscillations its 
approximation by CD,  
\begin{eqnarray} \label{pkdf} 
\delta_{1/t}(\ep) \approx \Delta _{1/t}(\ep) \,\, , \,  
\end{eqnarray}
seems reasonable. 

{\bf 6}.\, Returning to the general case \,$g(\ep)$\,, 
at sufficiently large observation time, when  
 \,$\langle n \rangle \gg 1$\,, 
 and EVs are dominated by QDF's tails, we can write 
\begin{eqnarray} \label{psc} 
\lm_t(\theta) \,\rightarrow\, 
\frac {\chi(\theta/\hbar)}t \,\, , \,\,\, 
\\   \nonumber  
\chi(x) \equiv \frac \hbar{\pi}  
 \int g(\ep)\, \frac {\cos{(\ep x)} -1}{\ep^2}  \,d\ep\, \, , \,\, 
\end{eqnarray}
and for propagator  (\ref{poe}) respectively   
\begin{eqnarray} \label{ppoe} 
S_{1/t}(t^\prime,\ep) \,\rightarrow 
\int e^{\,(t^\prime/t)\, \chi(x)}\, 
e^{\,i\ep x} \,\frac {dx}{2\pi}  \,\, .  \,\,    
\end{eqnarray}
Thus, according to (\ref{prope}) and (\ref{ppoe}), 
in most general case too the departure \,$\ef-\ef_0$\, 
of energy from its initial value becomes frozen 
with (quasi-) stationary distribution possessing finite width 
and characteristic function 
\,$\langle \exp{[ix\,(\ef-\ef_0)]} \rangle\, \rightarrow\, \exp{\,\chi(x)}$\,. 
 
All  that means that the remarkable property 
of summation of independent energy's  increments 
(discrepancies, uncertainties) like dependent ones 
survives when QDF's tails are cut off by factor \,$g(\ep)$\, 
(though may produce less tailed distributions than CD).
 
Naturally, time evolvent of such summation law for \,$\ef$\,'s increments 
looks as \,$\ef$\,'s fluctuations of flicker type, with infinitely 
propagating correlations.

{\bf 7}.\, At last, demostrate these correlations, 
in the meantime confining ourselves by two-time ones. 

Consider the fluctuation \,$\ep=\ef-\ef_0$\, as random process 
 \,$\ep(t)$\,. 
 Exploiting PKO in our GMA  similarly to use of KO in GRA, 
 introduce joint characteristic function 
 of \,$\ep(t)$\, and \,$\ep(t_0)$\, (с \,$0<t_0<t$\,) 
by expression 
\begin{eqnarray} \label{hfe}  
\langle \,e^{\,ix \ep(t)}\, e^{\,iy \ep(t_0)}\,\rangle 
= \,\,\,  \\  \nonumber \,\,\,  = 
 \int \!\!  d\ep\,\, e^{\,ix \ep}\, e^{\,(t-t_0)\kof_{1/t}} \, 
 e^{\,iy \ep}\,e^{\,t_0\kof_{1/t}} \, \delta(\ep) \,\, , \, 
 \end{eqnarray}
while operator exponentials (propagators) there we express  
like in (\ref{prope}) through integral operator (\ref{poe}). 
Besides, being interested in large time scales, 
we apply to (\ref{poe}) the approximation (\ref{ppoe}). 
This implies  
\begin{eqnarray} \label{ahfe}   
\langle \,e^{\,ix \ep(t)}\, e^{\,iy \ep(t_0)}\,\rangle 
 \, \rightarrow  \,  
e^{(1-t_0/t)\, \chi(x)} \, e^{(t_0/t)\,\chi(x+y)} \,\, . \,\,\,  
 \end{eqnarray}

It is not hard to deduce from here that  
\begin{eqnarray} \label{kk}   
\langle\, \ep(t)\,\ep(t_0)\, \rangle 
 \, = \, \frac {t_0}t \, \langle \ep^2 \rangle \,\, , \,\,\,  
 \end{eqnarray}
where 
the variance \,$\langle \ep^2 \rangle$\, must be taken from (\ref{dis}). 
This formula shows that correlation function of energy flucuations 
always has flicker type (slow, non-integrable) dependence  
on difference of its time arguments, 
and that its flicker behavior may be manyfold, from inverse proportionality 
to constancy, corresponding to quasi-static fluctuations.

\section{Statistical-mechanical appeoach to pseudo-kinetics}

\subsection{The Liouville equation and evolution of quantum state's 
probability distribution}

The formally exact description of our system with Hamiltonian (6), 
 \,$H=H_0+\Phi$\,, requires density matrix (DM)   \,$\vro$\, 
 and quantum Liouville equation 
(von Neumann equation) 
\begin{eqnarray} \label{le}
\frac {\partial \vro}{\partial t} = \lof\vro = 
(\lof_0 +\lof_\Phi)\,  \vro  \,\,
\end{eqnarray}
where \,$\lof$\, is Liouville super-operator whose action onto operators 
(``matrices'')  is defined by   
\begin{eqnarray} \label{lo} 
 \lof\vro = -\frac i\hbar\,[H,\vro] = 
 -\frac i\hbar\,(H\vro -\vro H)  \,\, \,  
\end{eqnarray} 
with  \,$\lof_0$\, and  \,$\lof_\Phi$\, representing two parts of \,$H$\,. 
  
As before, we will consider the system's DM and its evolution 
within the basis formed by eigen-states of  \,$H_0$\,, so that
the probability distribution of these states  \,$\rho$\,, 
which was under our interest, now is the DM's diagonal:\, \,$\rho_1 =\vro_{11}$\,. 
Sometimes we will temporarily identify \,$\rho$\, and matrix produced 
by turning all non-diagonal elements of \,$\vro$\, into zero, 
which may be written in the form \,$\rho = \dof\vro$\, with \,$\dof$\, being 
super-operator of this procedure. Obviously, it is projection (super-) operator, 
that is possessing property \,$\dof^2=\dof$\,. We introduce also 
the complementary projection (super-) operator 
\,$\nof= 1-\dof$\, and notice that \,$\nof\dof=\dof\nof =0$\,.  

We assume, as implicitly before, that the Hamiltonian  of interaction between 
system's parts, \,$\Phi$\,, the same as perturbation Hamiltonian, 
is purely non-diagonal, therefore operator identity   
\,$\dof\lof_\Phi \dof =0$\, takes place. Besides, taking into account 
one more operator identity \,$\lof_0 =\nof\lof_0$\,, we can write   
\begin{eqnarray} \label{lo+} 
 \lof = \nof\lof \nof + \nof\lof_\Phi\dof + \dof\lof_\Phi\nof    \,\, .\,  
\end{eqnarray} 

Further, let at  \,$t=0$\, the DM is purely diagonal, \,$\nof\vro(0)=0$\,, 
that is does not contain correlations between the basis states and 
is fully determined by their probability distribution \,$\rho (0)$\,.  
Then later  
\begin{eqnarray} \label{tev} 
\rho(t) = \sof(t)\,\rho(0)  \equiv  
\dof \,\exp{(t\lof)}\,\dof\,\rho(0)  \,\, . \,
\end{eqnarray} 
With the help of above identities and standard rules of  disentangling of 
operator exponentials the propagator \,$\sof(t)$\, is rewritable as 
\begin{eqnarray} \label{tpr} 
\sof(t) =\dof \,\, \overleftarrow{\exp} \,
[\int_0^t (\, e^{-t^\prime \nof\lof \nof}  \nof\lof_\Phi\dof
+ \,\,
\\   \nonumber  \,\, + \,\,
\dof\lof_\Phi\nof\,e^{\,t^\prime \nof\lof \nof} \,)\,dt^\prime \, ] \,\dof  \,\,  \,  
\end{eqnarray} 
with left arrow marking chronological ordering of exponentials. 
On the other hand, separating  diagonal and non-diagonal  parts of DM 
in the Liouville equation (\ref{le}) and applying (\ref{lo+}), 
it is not hard to derive for this propagator an integro-differential equation   
\begin{eqnarray} \label{dpr} 
\frac {\pa\sof(t) }{\pa t}  = \int_0^t \qof(t-t^\prime)\,  
\sof(t^\prime) \,dt^\prime \,\,  \,\,  
\end{eqnarray}  
with operator kernel   
\begin{eqnarray} \label{qo} 
\qof(t)  \equiv \, 
\dof\lof_\Phi\nof\,e^{\,t\,\nof\lof \nof} \,
 \nof\lof_\Phi\dof  \,\, . \,\, 
\end{eqnarray}

These exact formulas must be simplified, -  as far as the interaction 
 \,$\Phi$\,  allows to manage with lowest order of the quantum 
 non-stationary perturbation theory (QNPT), - by means of replacement 
\begin{eqnarray} \label{np} 
\nof\lof \nof \,\Rightarrow \, \nof\lof_0 \nof  \,\, , \,\,  
\end{eqnarray}
which will be named ``weak interection approximation'' (WIA).  
Using it in (\ref{qo}) and uncovering the super-operations, 
one can verify that   
\begin{eqnarray} \label{qo0} 
 (\qof(t)\,f)_1 \Rightarrow  
 \sum_2 \frac {2|\Phi_{12}|^2} {\hbar^2}\, 
 \cos \frac {\ef_{12}t}{\hbar}\,\, [\,f_2-f_1]  \,\, \,\,  
\end{eqnarray}  
with \,$\ef_{12}\equiv \ef_1-\ef_2$\,, i.e. 
\,$\qof(t) = d^2 t\kof_{1/t} /dt^2 $\,.

\subsection{From rigorous statistical mechanics to the pseudo-kinetics}

Let us consider the formula (\ref{tpr}). 
There the first of the two (super-) operators in the exponential 
creates inter-state correlations, i.e. non-diagonal 
matrix elements (ME) while the second annihilates them. 
At that, only those terms of the exponential's series expansion 
contribute to (\ref{tpr}) what include equal numbers of creations and 
annihilations to combine then into pairs.  

It is clear also that the  creation-annihilation pairs do follow  
one after another, with no time intersections, that is dividing the 
time among themselves and uncorrelated evolution. 
Nevertheless, any pair may occupy any part of available time, 
up to its whole value. Therefore in the framework of WIA 
it seems reasonable to neglect the exclusion of intersections 
and enumerate the pairs as if they were not feeling one another. 
In such manner (\ref{tpr}) yields approximation   
\begin{eqnarray} \label{ppr} 
\sof(t) \, \approx \, \sof_{1/t}(t) = 
\exp{(t\kof_{1/t})}  \,\equiv \,\, \,\,\,\,\,
\\  \nonumber  \equiv \,
\exp \iint_{t>t_1>t_2>0}  
\dof\lof_\Phi\nof\,e^{\,(t_1-t_2) \, \nof\lof_0\nof} \,
 \nof\lof_\Phi\dof  \,\,  \,\, , \,\,
\end{eqnarray} 
which, in view of (\ref{qo}) and (\ref{qo0}), is nothing but the 
golden mean approximation (GMA) with the pseudo-kietic operator 
(PKO) \,$\kof_{1/t}$\,.   

This derivation of GMA shows its tendency to somehow 
exaggerate a typical time 
given to transitions and  thus underestimate effects of uncertainty principle. 
Hence, it is interesting to see what things in the pseudo-kinetics 
may change when accounting for the time repulsion and competition between 
transitions.

\subsection{Mechanical corrections to pseudo-kinetics}

Next, consider equation  (\ref{dpr}) with kernel (\ref{qo0}) 
and apply to it the Laplace transform: 
\begin{eqnarray} \label{s+} 
\stf(z)\equiv \int_0^\infty e^{-zt}\,\sof(t) \,dt = 
 [ z - \qtf(z) ]^{-1}  \,\, ,  \,\,\,  
\end{eqnarray} 
where the tilde in place of hat decorates transforms instead of 
operator originals, and    
\begin{eqnarray} \label{lqo}  
(\qtf(z) \,f)_1  = \sum_2 \, \frac {2\pi}\hbar \, |\Phi_{12}|^2 \,
\Delta_z(\ef_{12})\, [\,f_2 - f_1] \,\, . \,\,\, 
\end{eqnarray}
It is visible that action of operator  \,$\qtf(z)$\,  
resembles that of PKO, but with another quasi-delta-function (QDF), 
in the form of the Cauchy distribution (CD) from (25) 
although with generally complex-valued ``width'' parameter  \,$z$\,. 
This QDF, like the old one in PKO, serves to improve kinetics of interaction 
by involving into it the uncertainty principle, 
but now making it more correctly (even maybe precisely). 
 
Let us consider spectrum of operator \,$\qtf(z)$\, 
as related to the interaction uncertainty in itself, 
that is to \,$\ef$\, dependency, similarly to previous section,   
reducing \,$\qtf(z)$\, to one-dimensional operator:   
\begin{eqnarray} \label{lqe} 
\qtf(z)\, f(\ef) = \int \Delta_z (\ef-\ef^\prime)\, 
\, \times\,\,\, \\   \nonumber   \times\,\, 
g(\ef -\ef^\prime)\,\,[\, f(\ef^\prime) - f(\ef)]\,\, d\ef^\prime \,\, .  \,\,    
\end{eqnarray}
This operator has the same eigen-functions (EF) 
but with different eigen-values (EV) 
\begin{eqnarray} \label{lsc} 
\wt{\lm}_z(\theta) = \int \Delta_z (\ep)\,  
g(\ep)\, [\,\cos{(\theta\ep/\hbar)} -1\,] \,d\ep\, \, , \,\,\, 
\end{eqnarray}
now complex if \,$z$\, is complex. The corresponding direct analogue 
of the pseudo-kinetical propagator kernel (20) is the following  kernel 
of the statistical-mechanical propagator \,$\sof(t)$\,:  
\begin{eqnarray} \label{pqe} 
S(t,\ep) = \int e^{\,i\theta\ep/\hbar}
\!\! \int _{-i\infty+0} ^{i\infty+0} \,
\frac {\exp{(tz)}}{z-\wt{\lm}_z(\theta) } \, 
\frac {dz}{2\pi i} \frac {d\theta}{2\pi\hbar}  \,\,   \,\,\,    
\end{eqnarray}
with \,$\ep=\ef-\ef^\prime $\,. For large enough time, the internal integral 
here is determined by small \,$z$\, region, and its asymptotic    
at \,$t/|\theta|\rightarrow\infty$\, can be calculated by rule 
\begin{eqnarray}     \label{rule} 
\lim_{t\rightarrow  \infty}
\int _{-i\infty+0} ^{i\infty+0} \,\dots\, \frac {dz}{2\pi i} \,
=\,  \lim_{z\rightarrow +0} \,\,\frac {z}{z-\wt{\lm}_z(\theta)} \,\, . \,\,\,    
\end{eqnarray}
Since  \,$\lim_{z\rightarrow +0} \wt{\lm}_{z}(\theta)/z = 
  \chi(\theta/\hbar) $\,, - with \,$\chi(x)$\, defined in (27), - one obtains    
\begin{eqnarray} \label{qork} 
S(\infty,\ep) = W(\infty,\ef_0+\ep) \rightarrow  \int 
\frac {\cos{(\ep x)}} {1 - \chi(x)} \,\frac {dx}{2\pi}  \,\, .  \,\,\,    
\end{eqnarray}
So, in rigorous statistical mechanics too the deviation \,$\ep=\ef-\ef_0$\, 
tends to a localized (quasi-) stationary distribution.  
But it  differs from that in pseudo-kinetics: now its 
characteristic function (CF) is  
\begin{eqnarray} \label{qhf} 
\langle \exp{[ix\,(\ef-\ef_0)]} \rangle\, \rightarrow\, 
 [\,1 - \chi(x)]^{-1} \, \, , \,\,\, 
\end{eqnarray}
instead of 
\,$\langle \exp{[ix\,(\ef-\ef_0)]} \rangle\, \rightarrow\,\exp \,\chi(x) $\, 
there. 

Comparing the limit distribution (\ref{qork}) with GMA result  (28) 
(at \,$t^\prime =t$\,), one can see that their variances and their tails at  
 large \,$\ep^2\gtrsim \langle \ep^2 \rangle$\, are coinciding, 
 and all their differences concentrate at  small \,$|\ep|\ll \ep_0$\, 
 being determined there by the asymptotic 
  \,$\chi(x \rightarrow\infty) = - \hbar g(0) |x|$\,.  
Under GMA this asymptotic leads to constancy of density of distribution 
in vicinity of \,$\ep=0$\,, but now, in  (\ref{qork}), it means 
a logarithmic divergency near \,$\ep=0$\,. Now   
\begin{eqnarray} \label{qrk} 
S(\infty,\ep) \approx \int 
\frac {\cos{(\ep x)}} {1 +\vep |x|} \,\frac {dx}{2\pi} 
\approx 
\frac 1{2\pi\vep} \, \ln\,\left(1+ \frac {2\vep^2}{\ep^2} \right) \,\, \,\,\,    
\end{eqnarray}
with  \,$\vep =\hbar g(0)$\,, when \,$|\ep|\ll \ep_0$\,. 
If  \,$g(\ep)=g=$\,const\, then this expression is valid anywhere, 
thus replacing CD. 

We see that statistical mechanics suggests noticable 
corrections to intuitive pseudo-kinetics, but seemingly non-principal ones. 
At last, compare in this respect two-time correlations.

\section{Flicker correlations} 

\subsection{Naive treatment of 
energy uncertainty correlation function}

Now consider two-time correlators of the interaction (enery) fluctuations  
\,$\ep(t)=\ef(t)-\ef_0$\, (deviations from start value \,$\ef_0$\,). 
For this purpose one should first introduce a definition of such 
correlators. Under statistical mechanical approach to pseudo-kinetics, 
most simple and seemingly natural definition is 
\begin{eqnarray} \label{hfe}  
\langle \,e^{\,ix \ep(t)}\, e^{\,iy \ep(t_0)}\,\rangle 
= \,\,\,  \\  \nonumber \,\,\,  = 
 \int \!\!  d\ep\,\, e^{\,ix \ep}\, \sof(t-t_0)\, 
   e^{\,iy \ep}\,\sof(t_0) \, \delta(\ep) \,\,  \,\, 
 \end{eqnarray}
with \,$0<t_0<t$\,. At not too small  \,$t_0\,\gg 1/g(0)$\, and 
 \,$t-t_0\,\gg 1/g(0)$\,  it yields 
\begin{eqnarray} \label{hfe+}   
\langle e^{\,ix \ep(t)} \,e^{\,iy \ep(t_0)}\rangle \,\rightarrow\, 
\frac 1{1-\chi(x)} \,\cdot\, \frac 1{1-\chi(x +y)} \, ,\, \,\,\,   
\end{eqnarray}
and, consequently (after differentiations over test variables \,$x,\,y$\, 
at point \,$x=y=0$\,),  
\begin{eqnarray} \label{kk+}   
\langle\, \ep(t)\,\ep(t_0)\, \rangle 
 \, \rightarrow \, \langle\, \ep^2\, \rangle  \,\, , \,\,\,  
 \end{eqnarray}
where variance on right-hand side is given by (23).

Thus, the correlation function (CF) of  
\,$\ep(t)$\, and \,$\ep(t_0)$\, occurs 
non-decaying at arbitrary growth of \,$t-t_0$\,, 
which corresponds to frozen  (quasi-static) fluctuations.

In pseudo-kinetics, similar definition is   
\begin{eqnarray} \label{hfe1}  
\langle e^{\,ix \ep(t)}\, e^{\,iy \ep(t_0)}\rangle \,=\,\,\,\,\, 
\,   \\  \nonumber  = \,
  \int \!\!  d\ep\,\, e^{ix \ep}\, \sof_{1/t_1}(t_1)\, e^{iy \ep} 
 \sof_{1/t_0}(t_0) \,  \delta(\ep)  \,\,  \,\,  
 \end{eqnarray}
with \,$t_1=t-t_0$\, and results in 
\begin{eqnarray} \label{hfe1+}  
\langle e^{\,ix \ep(t)} \, e^{\,iy \ep(t_0)}\rangle 
  \,\rightarrow\,   e^{ \,\chi(x)} \,e^{\,\chi(x +y)} \,\,  \,\,  
 \end{eqnarray}
and, evidently, in exactly the same 
asymptotically frozen  CF (\ref{kk+}). 

%---------------

 This agreement, nevertheless, is not too significant, 
 for quite similar CF behavior characterizes usual Gaussian 
 random walk (Brownian motion).  
But our random walk \,$\ep(t)=\ef(t)-\ef_0$\,, 
in opposite to the usual one, 
possesses the fast saturating dispersion (\ref{dis}) 
instead of infinitely growing one. 
In combination with (\ref{kk+}), it means  
(under not too small  \,$t_0 \gg \hbar/\ep_0$\,) that   
\begin{eqnarray} \label{d1}
\langle [\ep(t)-\ep(t_0)]^2 \rangle \,= 
\langle \ep^2(t)\rangle + 
\langle\ep^2(t_0)\rangle \, - 
\,\,\,\,\,  \\  \nonumber  \, - \,
\,2\, \langle \ep(t)\,\ep(t_0)\rangle   \, \rightarrow\, 0 \,\, , \,\,\,
\end{eqnarray} 
as if our walk was equally fast ``stopped''. 
This, however, looks very unnaturally, 
because the step \,$\ep(t)-\ep(t_0)$\, 
in principle must have the same numerical rights 
as the previous step  \,$\ep(t_0) - \ep(0)=\ep(t_0)$\,. 

Hence, in fact, our above ``Marcovian'' definition (\ref{hfe}) 
is incorrect from viewpoint of formally rigorous statistical mechanics 
and should be corrected with its help.

\subsection{Quantum correlation functions} 

In quantum statistical mechanics, 
because of non-commutativity of variables, 
their multi-time correlators can be introduced 
in many non-equivalent ways. Here, we accept the one 
which directly generalizes the unambiguous definition of 
multi-time correlators in classical statistical mechanics:\, 
if\, \,$O$\, are operators of quantum variables, 
then \,$n$\,-order joint statistical moment for them, 
taken at time points\, 
 \,$0\leq t_1 \leq t_2 \,\dots\,\leq t_n\leq t$\,, 
 is  
\begin{eqnarray} \label{qk}   
\langle\, \prod_{k=1}^n  O(t_k) \,\rangle  \,= \,
% \,\,\,\, \\ \nonumber  \,   \,=\, 
`\texttt{Tr}\,\{\, e^{(t-t_n)\lof} \,\jof_{O} \,e^{(t_n-t_{n-1})\lof} \,\dots\,
\,\,\,\,\, \\ \nonumber  \, 
 \,\dots\,  e^{(t_3-t_2)\lof} \,\jof_{O} \, e^{(t_2-t_{1})\lof} \,\jof_{O} \,  
 e^{\,t_{1}\lof} \,\, \vro(0) \,\}\,\,   , \,\,\,   
\end{eqnarray}
where \,\,$\jof_{O}$\, is super-operator 
of symmetrized, or Jordan, multiplication by operator  \,$O$\,:\, 
\,$
\jof_{O}\,A\,\equiv\, (O\,A + A\, O)/2\,\,  
$\,. 
 
Especially we are interested in diagonal \,$O$\,'s, 
that is commuting with  \,$H_0$\,, i.e. with  
the ``unperturbed energy'' (UE) \,$\ef$\,. 
For such \,$O$\,'s we have identity\, 
\[
\jof_{O} = \dof\jof_{O} \dof + \nof\jof_{O} \nof\,\,. 
\]
 Appplying it in (\ref{qk}) with two variables,\, 
\,$O(t_1)\Rightarrow A(t_0)$\, and  \,$O(t_2)\Rightarrow B(t)$\,,   
using also the decomposition (\ref{lo+}), 
propagator  (\ref{tpr}), 
WIA (\ref{np}),
properties of the trace \,$\texttt{Tr}$\,, 
and noticing that \,$[O,H_0]=0$\, implies 
 \,$[\jof_O\,,\nof\lof_0 \nof] =0$\,,  
 one obtains 
\begin{eqnarray} \label{qk2}   
\langle  B(t) \,A(t_0) \rangle = 
\texttt{Tr}\,[ B\,\sof(t-t_0)\, A \,\sof(t_0)\,+\,\,\,\,\,\,  
\\ \nonumber  \,+  
\int_{t_0}^t \!\! db \int^{t_0}_0 \!\! da\,\,  B\,\sof(t-b)\, 
\qof_A(b-a)\,  \sof(a)\, ] \,\rho(0) \,\, , \,\,\,  
\end{eqnarray}
where now\, \,$\texttt{Tr}\,\dots =\sum_1(\dots)_1$\,,\, 
\,$A$\, and \,$B$\, are mere operators of multiplication by 
functions \,$A_1$\, and \,$B_1$\, 
of indices ``1'' of unperturbed states, 
and new operator in spaces of such functions 
(similar to (\ref{qo})) has appeared, 
\begin{eqnarray}    \label{nop0} 
\qof_A(\tau) \equiv  \dof\lof_\Phi\nof \,
e^{\,\tau\,\nof\lof_0 \nof} \,\jof_A\, \nof\lof_\Phi\dof  \,\, . \,\,\,  
\end{eqnarray}

\subsection{Rigorous statistical-mechanical corrections to 
the energy correlation function}

In particular, for functions depending on the UE only,\, 
\,$f_1 = f(\ef_1)$\,,  operator (\ref{nop0}) reduces to   
\begin{eqnarray}    \label{nop}   
\qof_A(\tau) \,f(\ef) = \! \int g(\ef \!-\! \ef^\prime)\,\times \,\,\,\, 
\\  \nonumber  \times\, 
\cos{ \frac {(\ef \!-\! \ef^\prime)\,\tau}{\hbar}}\, 
 \, \frac {A(\ef) \! + \! A(\ef\pr)}{2}
 \,[ f(\ef^\prime) \! -\! f(\ef)]\, \frac {d\ef^\prime}{\pi\hbar}\, \, .  \,\,
\end{eqnarray}
If we take \,$B=A=\ef-\ef_0$\,, then (\ref{qk2}) with (\ref{nop}) 
becomes rigorous statistical-mechanical definition for CF 
 \,$\langle \ep(t) \,\ep(t_0)\rangle$\, of random walk \,$\ep(t)=\ef(t) -\ef_0$\,. 

Obviously, the first right-hand term in  (\ref{qk2}) 
looks as the definition  (\ref{hfe1}), 
i.e. is made as for Marcovian random processes, 
and its contribution (when \,$B=A=\ef-\ef_0$\,) must coincide with (\ref{kk}). 
It can be easy verified in the light of previous sections. 
Hence, clearly, the second term of (\ref{qk2}), which contains (\ref{nop}), 
represents non-Marcovian  correction to (\ref{kk+}).  
It can be exactly calculated when \,$B=A=\ef-\ef_0$\,, 
if paying attention to parities of the functions there 
and, besides, to automatic normalization 
  \,$\int S(\cdot,\ep)\, d\ep =1$\,. 
The result is  
\begin{eqnarray} \nonumber  % \label{smkk0}   
\langle\,\ep(t)\,\ep(t_0)\, \rangle\, =\,\langle\,\ep^2(t_0)\,\rangle \,+\,\,\,\, 
\\  \nonumber +\, 
\int_{t_0}^t \!\! db \int^{t_0}_0 \!\! da\,\, 
\int g(\ep)\,\cos{[\,\ep\,(b-a)/\hbar]}\,\, \frac {\ep^2}2\, 
\frac {d\ep}{\pi\hbar} \,\,  \,\,\, 
\end{eqnarray}
and finally  
\begin{eqnarray} \label{smkk}   
\langle \ep(t)\,\ep(t_0) \rangle  = \frac {\hbar}{2\pi}\,
[\wt{g}(0)- \wt{g}(t_0)+ \wt{g}(t_1)- \wt{g}(t)] \rightarrow \,\, \,\,\,  
\\  \nonumber \,\rightarrow \,
\frac 12\, \langle \ep^2(t_0) \rangle \,\, , \,\,\, 
\end{eqnarray}
where \,$t_1=t-t_0$\,, 
\[
\wt{g}(t) =\int g(\ep)\,\cos{(\ep t/\hbar)} \,d\ep\,\,,\, 
\]
and 
the arrow presumes that \,$t_0\,, t-t_0 \gg \hbar/\ep_0$\, 
(i.e. time intervals between energy measurements 
are much longer than the measurements themselves).

Thus, the main correction to the UE's CF 
is that its residual (non-decaying) 
value (at  \,$t_1\rightarrow\infty$\,)  
in fact is twice smaller than predicted in (\ref{kk+}). 

The meaning of this change becomes clear when including it 
into calculation of variance of UE's step 
\,$\ep(t)-\ep(t_0)$\,. Instead of (\ref{d1}), it yields   
\begin{eqnarray} \label{smd}
\langle [\ep(t)-\ep(t_0)]^2 \rangle \,=\, \frac {\hbar}\pi\, 
 \int g(\ep)\, d\ep \,\, .  \,\,  \,\,\,
\end{eqnarray} 
We see that now step \,$\ep(t)-\ep(t_0)$\, 
acquires the rights exactly equal to that 
of the preceeding step\,$\ep(t_0)-\ep(0)$\,.  

Accordingly, the flicker correlations do acquire 
a more live character:\, 
now they are decaying although by half only 
and independently on time separation of measurements. 
In other words, relaxation time of  the flicker fluctuations
has no definite value and may occur arbitrary large. 

Thus, once again statistical mechanics significantly 
corrects a ``hand-made'' pseudo-kinetics, but at the same time 
visually confirming validity and importance of its principal claims 
to canonical kinetics.

%%%%%%%%%%%%%%%%%%%%%%%%%%%%

\,\,\, 

\,\,\,

     *   *   *   *   *   *   *   *   *   *   *     
     
     \,\,\,

Up to now we were considering 
interactions in themselves, like ``Cheshire cat's smile''  
in absence of the cat. 
From now we want to go to investigation of
``the smiling cat as such'', i.e. from fluctuations of 
the unperturbed energy (UE) itself 
to that of actually observable variables 
connected to interaction-induced irreversible processes  
and their flicker and 1/f noises. 
First of all, to wandering of a particle in one or another 
medium (e.g. in phonon medium like in \cite{eph}).

\section{Particle's wandering in quantum medium} 

Next, our system will be a particle inside a medium with 
``free motion Hamiltonian'' (\ref{h00}) and interaction Hamiltonian  \,$\Phi$\,. 
We are interested in medium-enforced fluctuations of velocity \,$V=P/M$\, 
of the particle  and its resulting random displacements\, 
\,$R(t,t_0)=\int _{t_0}^t V(t\pr)\,dt\pr$\, which give rights to name it  
 ``Brownian'particle' (BP). Especially interesting are those statistical properties 
 of BP's wandering which can be characterized as flicker fluctuations 
 of its diffusivity (turning into that of its mobility when adding 
 an external force to its Hamiltonian \cite{p157,bk2,ufn,99,pufn,p1008}). 

Let us write full indexes of the unperturbed (interaction-free) states
as pairs formed by the velocity \,$V$\, and ciphers which now will  
be symbolical indexes of medium's states. 
Then full UE  \,$\ef$\, becomes a function\, 
\,$\ef_{V1} =E + \ef_1$\, with\, \,$E=E(V)=MV^2/2$\, 
and\, \,$\ef_1$\, being BP's and medium's energies.

\subsection{Pseudo-kinetics of particle in thermostat} 

Let \,$W(V,\ef)$\, denote  density of joint probability distribution 
of the velocity and UE, 
\begin{eqnarray} \label{rse} 
 W(V,\ef)=  \sum_1 \,\delta(\ef-\ef_{V1})\,\rho_{V1}\, \, . \,\,    
\end{eqnarray}
If \,$\rho_{V1} = \delta(\ef_{V1} -\Sigma)/\mc{N}(\Sigma)$\, is normalized 
uniform microcanonical distribution, 
then   
\[
 W(V,\ef)=\delta(\ef-\Sigma)\,\mc{N}_{th} (\ef-E(V))/\mc{N}(\Sigma)\,  
\] 
with  
\,$\mc{N}_{th} (\Sigma) = \sum_1\,\delta(\ef_{1} -\Sigma) $\, 
representing density of states of the medium (thermostat).  
As far as the latter is large enough,   
we can use general property of (infinite) microcanonical 
statistical ensembles \cite{ter}:  
\begin{eqnarray}  \label{tp} 
  \mc{N}_{th}(\ef-E)/\mc{N}_{th}(\ef) \,\Rightarrow\, e^{ -E/T}\,\, , \,\,\,      
\end{eqnarray}
where  \,$T$\, is temperature of the ensemble (thermostat) 
determined by equality 
\,$1/T \equiv d\ln{\mc{N}_{th}(\ef)}/d\ef$\,     
(and indifferent to \,$\ef$\,'s shifts after thermodynamic limit). 

Hence, in microcanonical equilibrium (in the sense of the uniformity)   
 \,$W(V,\ef) \Rightarrow \delta(\ef-\Sigma)\,W_0(V)$\,, 
where we introduced  
  \,$W_0(V) =\exp{(-E/T)} \,\mc{N}_{th}(\Sigma)/\mc{N}(\Sigma)$\, 
  which is (normalized) Maxwell's velocity distribution. 
 
In pseudo-kinetics, however, this is not true equilibrium, 
for it does not mean stationarity in respect to UE distribution. 
Therefore the related velocity Maxwellian  
also occurs non-stationary and thus formally non-equilibrium 
(unachievable as such as ending of evolution). 
Nevertheless, \,$W_0(V)$\, may be reasonable approximation 
for quasi-stationary ensembles. 

%-----------------

To consider actual statistics of BP's motion,
 including flicker correlations, 
it seems natural to constrict scope of the original PKO 
\,$\kof_{1/t}$\, (or \,$\qof(t)$\,, if starting from the Liouville equation) 
to functions of only  \,$V$\, and \,$\ef$\,.
This simplification requires to assume identical probabilities 
of medium's states with equal energies.
It is logistic ansatz least if the medium 
initially was statistically equilibrium. 
If so, then  \,$\rho$\, looks as   
\begin{eqnarray} \label{rse+} 
\rho_{V1} = \int \frac {\delta(E+\ef_{1}-\ef)} {\mc{N}_{th}(\ef-E)} \, 
W(V,\ef)\,d\ef  \,\, , \,\,\,    
\end{eqnarray}
where in view of above equalities we can write  
\begin{eqnarray} \nonumber  %   \label{dos}
 \mc{N}_{th}(\ef-E) =\mc{N}(\ef_0)\, e^{(\ef-\ef_0)/T}\,W_0(V) \, \, . \,    
\end{eqnarray}
Inserting these expressions into original PKO  
and then taking projection (\ref{rse}) of the result 
onto subspace of functions of the pair\, \,$X\equiv \{V,\ef\}$\,,  
one obtains   
\begin{eqnarray} \label{kose} 
\kof_{1/t}\, W(V,\ef) = \iint \delta_{1/t} (\ef-\ef^\prime)\, 
\, \times\,\,\,\,\, \\   \nonumber   \times\,\, 
\, w(X|X^\prime)\, \left[\,\frac {W(X^\prime)}{w_0(X\pr)}  - 
 \frac {W(X)}{w_0(X)}\, \right]\,\,dX\pr \,\, , \,\,\,    
\end{eqnarray}\,
where\, \,$W(X)\equiv W(V,\ef)$\,,  \,$X^\prime \equiv \{V^\prime,\ef^\prime \}$\,, 
\begin{eqnarray} \nonumber  %   \label{w0}
w_0(X) \equiv \frac 
{ \mc{N}_{th}(\ef-E(V) ) }{\mc{N}(\ef_0) } \Rightarrow  
e^{(\ef-\ef_0)/T}\,W_0(V) \, \, , \,\,\,  
\end{eqnarray}
and 
\begin{eqnarray}   \label{fse} 
w(V,\ef\,|\,V^\prime,\ef^\prime) \equiv \frac {2\pi}\hbar\,\sum_{1,2} \,
\delta(\ef -\ef_{V1})\,\, \times  \,\,\, \,  \\ \nonumber   \times\,\,  
\, |\Phi_{V1\,V^\prime 2}|^2 \,\, \delta(\ef^\prime -\ef_{V^\prime2})\, 
/\mc{N}(\ef_0)   \,\,  \,\,\,    
\end{eqnarray}
with\, \,$E =MV^2/2$\, and  \,$E^\prime =MV^{\prime\,2}/2$\,. 
The latter function obeys obvious symmetry 
\[
\, w(V,\ef|V\pr,\ef\pr) =w(V\pr,\ef\pr|V,\ef)\,\, . 
\]

%------------------

Clearly, general distribution \,$W_{st}(X)$\, 
satisfying \,$\kof_{1/t}\,W_{st}(X)=0$\,,  
that is stationary with respect to (\ref{kose}), 
is \,$W_{st}(X)=$\,const\,$\times\,w_0(X)$\,.  
It represents statistical equilibrium 
with equiprobability of all system's states, 
which, however, never can reaize, 
because can not be normalized on UE axis. 

This fact means that evolution governed by (\ref{kose})  
must lead again to ``frozen'' (infinitely slowing down) 
regime with such quasi-stationary distribution 
\,$W(t,X)$\, that  \,$\kof_{1/t}W(t,X)\propto W(t,X)/t$\, 
under \,$t\rightarrow\infty$\,. 
In other words, from spectral point of view, 
the PKO defined by (\ref{kose}) -(\ref{fse}) 
acts as if it had no zero eigen-value (EV) 
instead possessing a set of nearly zero EVs \,$\propto 1/t\rightarrow 0$\,   
with eigen-functions (EFs) which tend to degenerated stationary EFs 
of limit KO \,$\kof_0$\,.  

Undoubtedly, such ``glassy'' behavior of 
the joint velocity-UE pseudo-kinetic evolution 
must be inherited by marginal statistics of velocity and BP's path. 
To make it visual, let us consider two characteristic classes 
of possible ``transition density'' functions \,$w(X|X\pr)$\,.

\subsection{Energy shift-invriant interaction and  
marginal PKO of velocity}

First of two interesting classes of  \,$w(X|X\pr)$\, 
appears when variety (``spectrum'') of possible types of transitions  
keeps constant under shifts along UE axis 
while their density changes proprtionally to density of states:
\begin{eqnarray} \nonumber   \label{inv}
w(V,\ef +\ep|V\pr,\ef\pr+\ep) = e^{\,\ep/T}\, w(V,\ef|V\pr,\ef\pr)\,\, . \,\,\,  
\end{eqnarray}
This property takes place, in particular, in case of phonon medium \cite{eph}.  
Generally, it implies that 
\begin{eqnarray}   \label{ifse}
w(X|X^\prime)  = e^{\,(\ef\pr-\ef_0)/T}\, w(V,V\pr; \ef-\ef\pr)\,\, , \,\,\,
\end{eqnarray}
with asymmetric function of three arguments, such that 
\begin{eqnarray}   \label{as} 
w(V\pr,V;-\ep) =e^{-\ep/T}\,w(V,V\pr;\ep)\,\, , \,\,\,
\end{eqnarray}
and PKO is visually strictly invariant with 
respect to equal shifts of UE values. 

As the consequence, it becomes easy to reduce (like in \cite{last}) 
the joint velocity-UE PKO to marginal velocity's PKO 
for\,  \,$W(V) =\int W(V,\ef)\,d\ef$\,,    
merely by integrating (\ref{kose}) over UE. 
The result is 
\begin{eqnarray} \label{kos} 
\kof_{1/t}\, W(V) = \int  \! dV\pr \,\, 
\, \times\,\,\,\,\, \\   \nonumber   \times 
 \left[\, w_{1/t}(V,V\pr)\,\frac {W(V\pr)}{W_0(V\pr)}  - 
 w_{1/t}(V\pr,V)\,\frac {W(V)}{W_0(V)}\, \right]\,\,  \,\,  
\end{eqnarray}\,
with transition's density 
\begin{eqnarray}   \label{fs} 
w_{1/t} (V,V\pr) \equiv 
\int \delta_{1/t} (\ep)\,w(V,V\pr; \ep)\,d\ep\,\, . \,\,\,    
\end{eqnarray}
The latter, in contrast to (\ref{fse}), 
is non-symmetric function: 
\begin{eqnarray}   \label{ant} 
w_{1/t}^{ant}  (V,V\pr) \equiv \frac 12\, 
[w_{1/t}(V,V\pr) \! -\! w_{1/t}(V\pr,V)] \neq 0\,\,  \,\,\,     
\end{eqnarray}
at \,$V\neq V\pr$\, and \,$1/t\neq 0$\,. 

%-------------------------

In order to make these expressions less abstract, 
let us notice (as in \cite{last}) that function \,$w(V,V\pr;\ep)$\, 
by its definition must be proportional to  
(relative) medium's density of states at 
minimum of two (``left'' and ``right-hand'' in (\ref{ifse})) 
medium's energies,   
 \,$\Sigma = \ef\! -\! E(V)$\, 
and \,$\Sigma\pr = \ef\pr\! -\! E(V\pr)$\,.  
This physically clear statement implies  
\begin{eqnarray}    \label{wf} 
w(V,V\pr;\ep)  \,=\, w(V,V\pr|\, |E-\ep -E\pr|)\, 
\times  \,\,\,\,\,  \\ \nonumber  \,\times\,\, 
\exp{\left(-\frac {E-\ep +E\pr+\, |E-\ep -E\pr|}{2T} \right)} \,\,  \,    
\end{eqnarray}
(\,$\ep=\ef-\ef\pr$\,),  
where factor\,  
\,$ w(V\pr,V|\,\sigma)$\, 
represents 
(chances or probabilities of) 
transitions as such,
with\,  
\[
 \sigma \equiv |E-\ep -E\pr|  = |\Sigma -\Sigma\pr| \, 
 \]
 being energy given up or taken out by medium in separate transition. 
This formula expresses that, naturally, 
chances, or probabilities, of particular transitions  
can depend on the UE ``misclosure'' \,$\ep$\, only indirectly 
through the medium energy changes \,$\sigma$\,. 

Thus, combining  (\ref{fs}) and  (\ref{wf}), we can write
\begin{eqnarray}   \label{fs+} 
w_{1/t} (V,V\pr) = e^{-E\pr/T} 
\int_0^\infty \! d\sigma\,w(V,V\pr|\sigma) \,\, 
\times \,\,\,\,\,\,\,  \\  \nonumber \times\,\, 
[\,\delta_{1/t} (E-E\pr +\sigma) + e^{-\sigma/T}\, 
\delta_{1/t} (E-E\pr -\sigma)\,]\,\, . \,\,    
\end{eqnarray}
At that, \,$w(V,V\pr|\,\sigma) $\, is symmetric function of the 
velocity arguments, while its actual dependence on \,$\sigma$\, 
reflects one or another concrete variety 
of  excitations of the medium and energy quanta 
what can be irradiated or absorbed by it. 
The PKO from \cite{eph} gives such the example.

\subsection{Non-stationarity, glass-likeness and non-self-adjointness 
of (equilibrium) particle's wandering pseudo-kinetics} 

Because of non-zero anti-symmetric component of \,$w_{1/t} (V,V\pr)$\,, 
(\ref{ant}),    
\[
 \kof_{1/t}\, W_0(V) =2\int w_{1/t}^{ant}(V,V\pr) \,dV\pr  \neq 0\,\, , \, 
\]
thus confirming that Maxwellian velocity distribution 
is not stationary one. 
At the same time, with no doubts, it adequately describes 
statistical equilibrium between BP and medium. 
This contradiction prompts that in fact PKO (\ref{kos}) has no 
stationary distribution at all, i.e. has no zero EV! 

To prove this, it is sufficient to notice 
that existence of such distribution, \,$W_{st}(V)$\,, 
would mean that the joint PKO (\ref{kose}), in addition to 
above mentioned \,$W_{st}(V,\ef)\propto w_0(X)$\,, 
has one more strictly stationary distribution 
 \,$W_{st}(V,\ef)\propto W_{st}(V)$\,. 
But such duplication is incompatible with  
UE's tendency to equipartitioning over system (micro-) states. 
 
For simple illustration, consider special factorized case 
\,$w(V,V\pr|\sigma) = w(V)\,g(\sigma)\,w(V\pr)$\,.  
Then from the stationarity equation 
 \,$\kof_{1/t}\, W_{st}(V) =0$\, 
and (\ref{fs+}) (with bounded \,$\sigma$\,'s) 
it would follow that 
\[
 W_{st}(V) \rightarrow \texttt{const}\, 
 \,\,\,\, \texttt{at}\,\,\,\,\, 
 E(V)\,  \rightarrow \infty\,\, , \,  
\]
with\, 
\,const\,$=\int w(V\pr)\,W_{st}(V\pr)\,dV\pr / \int w(V\pr)\,dV\pr$\,,  
thus demostrating impossibility of sensible solution.  

Nevertheless  PKO (\ref{kos}) must ensure stable quasi-stationary 
(``frozen'') asymptotics of  \,$W(t,V)$\,'s evolution. 
This means that (\ref{kos}), like (\ref{kose}),  replaces zero by  
small EVs \,$\propto 1/t$\,.   
 Alternatively one may imagine more exotic situation 
 with absence of any EVs. 
 
Simultaneously, the conjugated, in the Sturm-Liouville sense, i.e. transposed, 
operator \,$\kof_{1/t}^\dag$\,, acting by  
\[
\kof_{1/t}^\dag F(V) = W _0^{-1}(V) 
\int w_{1/t}(V\pr,V)\, [\, F(V\pr)-F(V)]\, dV\pr\, ,  
\]
certainly has zero EV with EF \,$F(V) =$\,const\,, 
which merely reflects the fact that PKO always conserves probability:\, 
\,$\int \! dV\, \kof_{1/t}\,\dots =0$\,. 
Hence, anyway our PKO is essentially non-self-adjoint 
(can not be symmetrized by a non-singular transformation). 

These circumstances were not fully realized in \cite{eph},  
although suggested by exact formulas of \cite{eph}.   
But in future just they may direct practical approximations 
of (pseudo-) kinetics of various interactions. 
For instance, similar to \cite{last}, where we have managed 
withot EVs and EFs in mind.  
Anyway our considerations develope understanding of that 
real kinetics, ruled by mechanics, always is more or less 
``glassy'', ``smeared'', ``jaming'', ``flickering'' and so on. 

%----------------------

\subsection{Case of energy shift-non-invriant interaction, 
wavelet eigen-functions, and tau-approximation}

Another interesting class of interaction density, 
or interaction intensity, functions  \,$w(X|X\pr)$\, 
is characterized by their indifference   
to density of medium states, so that 
\[
w(V,\ef +\ep|V\pr,\ef\pr+\ep) =  w(V,\ef|V\pr,\ef\pr) 
= w(V,V\pr; \ef\! -\! \ef\pr) \,\,  \,  
\] 
with \,$w(V,V\pr;\ep) =w(V\pr,V;-\ep)$\,. 
Then the joint velocity-UE PKO can be written as  product  
\[
 \kof_{1/t} = \kof^-_{1/t}\, \eta^{-1}(\ef-\ef_0)\,\, , \,  
\]
where\, \,$ \eta(\ep) \,\equiv \, e^{\,\ep/T}$\, and 
\begin{eqnarray} \label{kose+} 
\kof^- _{1/t}\, W(X) = \iint \delta_{1/t} (\ef\! -\! \ef^\prime)\, 
\, \times\,\,\,\, \\   \nonumber   \times\,\, w(V,V\pr; \ef\! -\! \ef\pr)\, 
 \left[\,\frac {W(X^\prime)}{W_0(V\pr)}  - 
 \frac {W(X)}{W_0(V)}\, \right]\,\,dX\pr \,\, . \,\,\,    
\end{eqnarray}

In opposite to previous case,  
structure of this PKO prevents UE's exclusion from consideration.  
But it allows another simplifications. 

By definion of operator \,$\kof^-_{1/t}$\, it is 
of difference type with respect to UE.  
Therefore, clearly, if \,$\kof_{1/t}$\, has EF 
\,$W_\lm(V,\ef) =w_0(X)\,\Psi_\lm(V,\ef)$\, 
with EV \,$\lm$\,, then UE-shifted function \,$W_\lm(V,\ef+\ep)$\,
is \,$\kof_{1/t}$\,'s EF with exponentially scaled 
EV \,$\lm\,\eta(\ep)=\lm\,\exp{(\ep/T)}$\,. 
Hence, such EF, if it exists, is a ``wavelet'' function. 

It is interesting picture, but we confine ourselves 
by its specific model example highlighting how slow UE fluctuations do ``modulate'' 
rate (spectral power) of faster BP velocity fluctuations. 

%---------------

To formulate the model, notice that 
in the GRA limit function (\ref{fse}) retains two of four arguments only, 
 \,$w(V,\ef|V\pr,\ef\pr) \Rightarrow \,w(V|V\pr)= w(V,\ef_0|V\pr,\ef_0)$\,, 
and PKO action reduces to    
\begin{eqnarray}  \nonumber  \label{ko} 
\kof_0 W(V) = \int  
%\, \times\,\,\,\,\, \\   \nonumber   \times\,\, 
w(V|V\pr)\, \left[\,\frac {W(V^\prime)}{W_0(V\pr)}  - 
 \frac {W(V)}{W_0(V)}\, \right]\,dV\pr \,\, \,\,\,\,     
\end{eqnarray}\,
with KO \,$\kof_0$\, from conventional Marcovian kinetics of particle in thermostat. 
It formally simplifies as much as possible, if
one attracts well known ansayz sometimes named ``tau-approximation'':\, 
\[
w(V,V\pr) \, \Rightarrow \, g\,W_0(V)\,W_0(V\pr)\,\, , \,\,\, 
\]
so that\, \,$\kof_0\, \Rightarrow \,  g\,(\wh{\Pi} -1) $\,,\,    
where \,$\wh{\Pi}$\, is projection operator acting onto 
velocity argument as follows,  
\[
 \wh{\Pi}\,F(V)\,\equiv\, W_0(V) \int F(V\pr) \,dV\pr\,\, . \,\,
\]

Let us transfer this caricature but useful model 
to the PKO by replacement 
\begin{eqnarray} \label{tau} 
w(V,\ef|V\pr,\ef\pr) \, \Rightarrow \, g(\ef-\ef\pr) \,
W_0(V)\,W_0(V\pr) \,\,  \,\,\,\,     
\end{eqnarray}\,
with \,$g(\ep) $\, being an even function 
(thus, as before, assuming UE shift-invariance of 
 intensity of transitions along with its insensitiveness to 
 density of states).  
 Then   
\begin{eqnarray} \label{kose0} 
\kof^-_{1/t} \,=\, \wh{K}_{1/t} \, \wh{\Pi} \, - 
\, g_t \, (1-\wh{\Pi}) \,\, , \,\,\,    
\end{eqnarray}
where \,$\wh{K}_{1/t}$\, is new designation for the operator (\ref{koe}), 
and \,$g_t = \int \delta_{1/t}(\ep)\,g(\ep)\,d\ep $\,.  

One can see that  second term of (\ref{kose0}) is responsible for fast 
relaxation of velocity distribution to vicinity of Maxwellian equilibrium,  
while first term for slow fluctuations in UE, then transformed, via 
factor  \,$1/\eta(\ep) =\exp{(-\ep/T)}$\,, to such  
in BP's wandering rate, i.e. its diffusivity. 

Factor  \,$1/\eta(\ep)$\,, besides, causes asymmetry of UE fluctuations. 
It was neglected above in the PKO (\ref{koe}) 
but now will be taken into account by introducing 
(normalized to unit) distribution    
\begin{eqnarray} \label{nre} 
W_T(t,\ep; \ep\pr ) \,\equiv\, \exp{(t \wh{K}_{1/t}\,e^{-\ep/T})}\,\, 
\delta(\ep-\ep\pr )\,\, , \,\,\,
\end{eqnarray}
which  at \,$T=\infty$\, turns to the above exploited one:\, 
 \,$W_\infty(t,\ep ; \ep\pr )= 
  W(t,\ep-\ep\pr )\equiv S_{1/t}(t,\ep-\ep\pr )$\,. 

Keeping in mind large time intervals, 
we will replace  \,$g_t$\, by \,$g\equiv g_\infty =g(0)$\, 
and  (\ref{nre}) by quasi-stationary limit 
\,$W_T(\ep ; \ep\pr )= W_T(\infty,\ep ; \ep\pr )$\,. 
The latter can be represented, - with help of function (\ref{psc}) 
and symbol \,$\nabla_x$\, for differentiation over \,$x$\,, - 
in the form   
\begin{eqnarray} \label{npor} 
W_T(\ep ; \ep\pr ) \,=\, \exp{[-\chi(\sqrt{-\nabla^2_\ep})\,e^{-\ep/T}\,]}\,   
\delta(\ep-\ep\pr )\,\,  \,\,\,\,
\end{eqnarray}
underlying that it is distribution of ``aborted'', 
stronly non-Gaussian and non-uniform, UE's diffusion. 
Of course, it is localized near \,$\ep\pr $\,, 
moreover, in such way that    
\begin{eqnarray} \label{etasm}
  \int \eta^{\,s}(\ep)\, W_T(\ep|\ep\pr )\, d\ep \,<\, \infty\,\, \,\,\,\,  
\end{eqnarray}
for integer \,$s\geq 0$\,. It can be deduced from consideration 
of eigen-value problem 
\begin{eqnarray}  \nonumber  %  \label{evp} 
-\chi(\sqrt{-\nabla^2_\ep})\,\Psi_\lm(\ep) = 
\lm\, e^{\,\ep/T}\,\Psi_\lm(\ep) \,\, , \,\,\,\,
\end{eqnarray}
with wavelet-like EFs \,$\,\Psi_\lm(\ep) $\,, 
and \,$W_T(\ep|\ep\pr )$\,'s expansion over them, 
but we here merely suppose (\ref{etasm}) satisfied at least for \,$s\leq 3$\,.  

Just formulated model is sufficient for demonstration 
how principal uncertainty of UE fully materializes in kinetics of BP.

\subsection{Particle's velocity correlation function and diffusivity} 

We introduce correlation function (CF) of velocity in analogy with (\ref{hfe1}):  
\begin{eqnarray} \label{okfs}  
C_2(\tau;t)\equiv \langle V(t+\tau)\, V(t)\rangle \,=\, 
\,\,\,\, \\  \nonumber  \, = \,  
\iint \!\!  d\ep\,dV\,\,V\,e^{\,\tau\, \kof_{1/\tau}}\,V \, 
e^{\,t\,\kof_{1/t}} \,W_0(V)\,\delta(\ep)\,\, . \,\,\,  
 \end{eqnarray}
Since initial velocity distribution is chosen Maxwellian, 
we speak about equilibrium fluctuations. 
For brevity and simplicity we will treat $\,V$\, as projection 
of velocity vector onto fixed direction. 

Notice that (\ref{kose0}) implies decomposition  
\begin{eqnarray} \label{raz}  
e^{\,t\,\kof_{1/t}} \,=\, \wh{\Pi}\,e^{\,t\wh{K}_{1/t} \eta^{-1}}\,+\, 
(1- \wh{\Pi})\, e^{-t g\eta^{-1}} \,\, . \,\,\,\,  
 \end{eqnarray}
At interval \,$(0,t)$\, in (\ref{okfs}) first term of this decomposition only
does work, but after multiplying by \,$V$\, its second term only, so that  
\begin{eqnarray} \label{kfs}  
C_2(\tau ;t)= V_0^2 \int \!\!  d\ep\,\, e^{-\tau g \eta^{-1}}   
e^{\,t \wh{K}_{1/t}\,\eta^{-1}} \,\delta(\ep)\,\rightarrow\,  \,\,\,\,  
\\ \nonumber \,\rightarrow\, 
 V_0^2 \int e^{-\tau g \eta^{-1}}\, W_T(\ep|0)\, d\ep\,\, , \,\,\,  
 \end{eqnarray}
where\, \,$V_0^2\equiv \int V^2\,W_0(V)\,dV =T/M$\, 
is equilibrium variance of velocity. 
Clearly, its CF decays with relaxation time  \,$\eta/g = \exp{(\ep/T)}/g$\, 
which appears as effectively random since dependient on variable \,$\ep$\,. 
 At that, inequality (\ref{etasm}) guarantees that BP's 
 diffusion coefficient (diffusivity)  
\begin{eqnarray} \label{kd}  
\df \equiv \int_0^\infty  C_2(\tau ;\infty) \,d\tau = 
\, \frac {V_0^2}{g} \int  \eta(\ep)\,W_T(\ep|0)\,d\ep\,\,  \,\,\,\,  
\end{eqnarray}
takes a finite value:\, \,$\df < \infty $\,. 
In such case, fluctuations of diffusivity (BP's diffusion rate), 
relative to its average \,$\df$\,, become interesting question.

\subsection{Flicker fluctuations of diffusion rate} 

Like the diffusion  coefficient (DC) is connected to quadratic correlation 
of velocity fluctuations, 
a correlation function of diffusion  coefficient's fluctuations 
(DC CF) naturally connects to fourth-order velocity correlations. 
More preccisely, to fourth-order cumulant 
\begin{eqnarray} \label{k4}  
C_4(\tau\pr,t,\tau;t_0) \equiv 
\langle \,V(t_0+\tau + t + \tau\pr)\,\,\times \, 
\,\,\,\, \\  \nonumber  \, \times \,
V(t_0+\tau + t)\, V(t_0+\tau)\,V(t_0)\,\rangle \,-\,  
\,\,\,\, \\  \nonumber  \, - \,
C_2(\tau\pr;t_0+\tau+t)\, C_2(\tau ;t_0)\,\,-\,\dots\,\,\, , \,\,\,  
\end{eqnarray}
where dots are replacement of two more products of two CFs 
representing two time-crossed parings between four multipliers. 
We want to know how integral over \,$\tau,\,\tau\pr$\, and  \,$t$\, 
in this expression do behave when the total time coverage 
of correlations,  \,$\tau + t + \tau\pr$\,, is infinitely growing. 
If the integral  also grows to infinity, 
then BP's wandering passes as if its DC was undergoing flicker fluctuations.   
An effective CF of these fluctuations logically \cite{p157,bk1,ufn,99} 
can be expressed through the fourth cumulant by  
\begin{eqnarray} \label{kfkd}  
K_\df(\theta) = \iint_{ \tau+\tau\pr <\theta}
C_4(\tau\pr,\theta- \tau-\tau\pr,\tau;t_0) \,d\tau\, d\tau\pr\,\,  \,\,\,  
 \end{eqnarray}
(increasing reference point \,$t_0$\, for stationarity). 
 
To consider this question, let us define the fourth-order statistical moment 
in  (\ref{k4}) similarly to (\ref{okfs}) by formula
\begin{eqnarray} \label{ok4}  
\langle \cdot\,\cdot\,\cdot\,\cdot \rangle \equiv  
\iint \!\!  d\ep\,dV\,\,V\,e^{\,\tau\pr\, \kof_{1/\tau\pr}}\,V \, 
\times \,\,\,\,\, \\   \nonumber \times\,\, 
e^{\,t\,\kof_{1/t}} \,V\, e^{\,\tau\, \kof_{1/\tau}}\,V \,
e^{\,t_0\,\kof_{1/t_0}} \,W_0(V)\,\delta(\ep)\,\, . \,\,\,  
 \end{eqnarray}
Using again decomposition (\ref{raz}) 
at intervals \,$(0,t_0)$\,, \,$\tau$\,, \,$t$\, and \,$\tau\pr$\,, 
and then performing integration over velocity, we obtain 
\begin{eqnarray} \label{k4+}  
\langle \cdot\,\cdot\,\cdot\,\cdot \rangle = V_0^4 \int \!\!  d\ep\,\,
 e^{-\tau\pr g \eta^{-1}}\,\, \times\, 
\,\,\,\,\, \\   \nonumber \times\,\,
[\,2\,e^{-t g\eta^{-1}} + e^{\,t\wh{K}_{1/t} \eta^{-1}} ]\, 
 e^{-\tau g \eta^{-1}}\,W_T(t_0,\ep|0)\,\, . \,\,\,  
 \end{eqnarray}
In view of (\ref{etasm}) integration of the first term in square brackets 
here over \,$\tau,\,\tau\pr$\, and \,$t$\, gives finite result, 
as well as integration of the dots in  (\ref{k4}). 
Hence, these contributions are insignificant for DC CF (\ref{kfkd}). 
What is for contribution from the difference of two other terms, 
it can be written (making \,$t_0 $\, large enough) as 
\begin{eqnarray} \label{pkfkd}  
K_\df(\theta) \,\rightarrow\, V_0^4 \iint \!\!  d\ep\,d\ep\pr   
\iint \!\! d\tau\,d\tau\pr \,\,  
e^{-\tau\pr g/\eta(\ep)}  \times\,\,\,  
\,\,\, \\   \nonumber \times\,
[\,W_T(t,\ep;\ep\pr ) - W_T(\ep;0) ]\, 
e^{-\tau g/\eta(\ep\pr )}\, W_T(\ep\pr ;0)\,\, , \,\,\,
\end{eqnarray}
where\, \,$t\equiv \theta- \tau-\tau\pr$\, and\, \,$\tau +\tau\pr <\theta$\,. 
This expression makes it clear that DC CF, together with 
 \,$W_T(t,\ep;\ep\pr )$\,, possesses 
 frozen asymptotics tending at large  \,$\theta \gg 1/g$\, 
to constant   
\begin{eqnarray} \label{akfkd}  
K_\df(\infty) \,= \iint \wt{\df}(\ep)\,  
[\,W_T(\ep;\ep\pr ) - W_T(\ep;0) ]\,\times\,  
\,\,\, \\   \nonumber \times\,
\wt{\df}(\ep\pr )\, W_T(\ep\pr ;0)\,d\ep\,d\ep\pr   \,\, , \,\,\,
\end{eqnarray}
where\,  \,$\widetilde{\df}(\ep) \,\equiv\, (V_0^2/g)\, \exp{(\ep/T)}$\,  
plays role of random diffusion rate varied by UE fluctuations.  

In case of weak interaction and ``rare BP-medium collisions'', 
in the sense of\, \,$\hbar g\ll\, T,\,\ep_0$\,, 
one can write approximately 
\[
\df \approx \frac {V_0^2}g\,\,, \,\,\,\, 
\wt{\df}(\ep) \approx \frac {V_0^2}g\, \left(\,1+ \frac \ep T \right)\,\, 
\]
and\, 
\,$W_T(\ep;\ep\pr) \approx W(\infty,\ep-\ep\pr)$\,,\, 
and come to estimate
\[
 K_\df(\infty) \,\approx\, \frac {\df^2\,\sigma^2(\infty)}{T^2}  
 \sim \frac {\ep_0}{\hbar g}\,  \left(\frac \hbar M \right)^2 
 \, \gtrsim\,  \left(\frac \hbar M \right)^2\,\, . \,\, 
\]

The constancy of this DC CF asymptotics 
naturally reproduces that of the CF (\ref{kk}) of UE itself. 
But, in principle, it may get essential corrections, 
like (\ref{kk}) was changed to (\ref{smkk}), 
in the framework of statistical-mechanical improvements  
of pseudo-kinetics.

\subsection{Towards full statistics of particle wandering} 

Instead of direct calculation of DC CF, possibly, 
it would be more comfortable to obtain it by  
considering whole statistics of BP's wandering, 
for instance, for beginning, characteristic function 
of BP's path,   
\[
\Xi(t-t_0,i\vk;t_0)=
\langle \,\exp{[\,i\vk\,R(t,t_0)]}\,\rangle \,\, . 
\]
Then once again we have to give definition 
of the angle brackets. 
Avoiding  refinements and fantasies  
of theries of quantum measourements \cite{men} 
(all the more their scholasticism) 
we will apply the recipe \cite{izm} 
which directly generalizes (\ref{qk})  
to ``continuous'' (i.e. arbitrary frequent) observations:  
\begin{eqnarray} \label{qhf}   
\langle \,\exp{[\,i\int_{0}^t \vk(t\pr)\, V(t\pr)\, dt\pr]}\,\rangle \,=\, 
\,\,\,\,\, \\ \nonumber \,=\, 
\texttt{Tr}\,\, \overleftarrow{\exp}\, 
\left\{\int_{0}^t [\,i\vk(t\pr)\,\jof_V + \lof\,]\,dt\pr \right\}\, 
\, \vro(0) \,\, . \,\,\,\,
\end{eqnarray}
Or we may consider ``correlation-characteristic function''   
\begin{eqnarray} \label{qhf+}  
C(t,i\vk) \equiv 
\langle\, V(t)\,\exp{[\,i\vk\int_{0}^t V(t\pr)\, dt\pr]}\,V(0)\,\rangle \,=\, 
\,\,\,\,\, \\ \nonumber \,=\, 
\texttt{Tr}\,\,V\, \exp{[\,t\,(i\vk \jof_V + \lof)]}\,\jof_V\,\vro(0) \,\, . \,\,\,\,
\end{eqnarray}
Most important task then is careful analysis 
of time dependencies with their ``non-Marcovian'' peculiarities 
caused by UE uncertainty 
(though, at that, other dependencies of imteractions' matrix elements  
and statistical ensembles allow rough approximations). 

For the case of energy shift-inavariant interaction 
(i.e. transitions' density proportional to states' density),
pointed out in Sections VI.B-VI.C, 
attempt of such analysis was made recently in \cite{last}. 
It verified inevitability of presence of flicker correlations 
and fluctuations in actual wandering statistics, 
qualitatively differing it from ideal Brownian motion, 
and gave them original quantitative estimate.
Or, to be precise, sooner that of their lower bounds, 
thus stimulating new approaches. 

\,\,\,

\section{Conclusion} 

Physical system being able to generate random events 
is unable to give them definite probability, or chance, 
if this system does not ke-averagable ep in memory their past numbers. 
Therefore their number per unit time in no way is stimulated to tend 
with time to a certain limit  
and constantly tries arbitrary long time-non-averagable 
deviations from expectations, i.e. flicker or  1/f type, flucuations. 
This trivial reasoning is regularly confirmed by experiments 
but hardly find place in scientific minds and 
stays in radical disagreement with unwritten dogmas 
and lexicon of physical-chemical kinetics. 

We have noticed that this contradiction 
can arise from disrespect of theoreticians for 
principle of uncertainty of energy and frequency 
in events on finite time arena. 
Kinetics neglects it when appeals to ``Fermi golden rule''. 
But quite elementary mathematics has revealed 
that very small uncertainties 
 \,$\sim \hbar/t$\, of energy in such events 
 as quantum transitions are summating with observation time \,$t$\ 
in proportion to their number \,$\sim gt$\, 
up to non-decreasing uncertainty \,$\gtrsim \hbar g$\, 
in energies of system states involved to the transtions.  
This stochastic process does not obey the ``law of large numbers'', 
so that the non-decreasing uncertainty transforms to 
flicker fluctuations in intermediate energies 
and pprobabilities of transitions 
and all other related quantities. 

This surprise of the theory appears already within  
own frames of kinetics if only it refuses  
the golden rule, as inconsistent with rigotorous 
statistical mechanics, and thus turns 
to  ``pseudo-kinetics''. 

Pseudo-kinetics prompts where theory hides sources 
of flicker noise and suggests its preliminary description, 
maybe rough but useful, because it surely finds 
qualitative confirmations and quantitative improvemens 
in statistical mechanics, to an extent of its mathematical 
development. 
 
We have demonstrated the said 
by relatively simple example on flicker fluctuations 
of transitions' energies. 

Then, we have shown that pseudo-kinetics 
unambiguously predicts flicker fluctuations 
in such practically observable physical quantity 
as rate (coefficient) of diffusion (wandering) 
of particle interacting with medium 
(equilibrium thermostat). 

Considering such effects of time-energy uncertainty, 
in pseudo-kinetics and in equations of 
statistical mechanics, we showed also that 
other aspects of quantum interactions and transitions 
can be simplified even to primitive models 
without loss of flicker noise. 

It, nevertheless, leaves non-standard 
mathematical difficulties to be resolved  
in future. 
Now, more important thing is that we already 
made several steps to first-principle 
microscopic theory of thermal flicker noise 
and determined aims of next necessary steps.

%%%%%%%%%%%%%%%%%%%%%%%%%%%%

\,\,\,

\end{document}